\documentclass[onecolumn]{jpsj3}
\usepackage{txfonts}

\title{Two Fermi surface states and two $T_{\rm c}$-rising mechanisms 
revealed by transport properties in $R$FeP$_{1-x}$As$_x$O$_{0.9}$F$_{0.1}$ 
($R$=La, Pr and Nd)}

\author{\name{Shigeki \surname{Miyasaka}}$^1$$\thanks{E-mail: miyasaka@phys.sci.osaka-u.ac.jp}$, 
\name{Akira \surname{Takemori}}$^1$, 
\name{Tatsuya \surname{Kobayashi}}$^1$, 
\name{Shinnosuke \surname{Suzuki}}$^1$, 
\name{Satoshi \surname{Saijo}}$^1$, 
\name{Setsuko \surname{Tajima}}$^1$ 
}

\inst{$^1$Department of Physics, Graduate School of Science, 
Osaka University, Toyonaka, Osaka 560-0043} 

\abst{We demonstrate the relation between critical temperature $T_{\rm c}$ 
and transport properties in $R$FeP$_{1-x}$As$_x$O$_{0.9}$F$_{0.1}$ 
($R$=La, Pr and Nd). 
$T_{\rm c}$ and resistivity power-law exponent $n$ form 
a universal line on the $T_{\rm c}$ vs. $n$ plane 
for all the $R$-systems with $x$$<$0.6$\sim$0.8, 
indicating that $T_{\rm c}$ increases with bosonic fluctuation. 
Transport properties show anomalies suggesting 
a change of Fermi surfaces around $x$=0.6$\sim$0.8. 
Above $x$=0.6$\sim$0.8, $T_{\rm c}$ and $n$ approach the second 
$T_{\rm c}$-$n$ line for the higher $T_{\rm c}$ systems. 
A further increase of $T_{\rm c}$ above $x$=0.6$\sim$0.8 
indicates the presence 
of an additional $T_{\rm c}$-rising mechanism in this system.}

\kword{superconductivity, iron-based superconductor, transport properties}

\begin{document}
\maketitle

\section{Introduction}

Since the discovery of superconductivity (SC) 
in iron pnictides~\cite{Kamihara}, 
a lot of experimental and theoretical efforts have been paid 
to find key parameters for determining high critical 
temperature $T_{\rm c}$ in this system. 
The pioneering work by Lee $et$ $al$. demonstrated 
that the crystal structure, particularly the bond angle of (As,P)-Fe-(As,P) 
is strongly correlated with $T_{\rm c}$~\cite{Lee}. 
However, further experiments have shown that 
Lees' conclusion is not applicable for all the iron based 
superconducting systems. 
Another parameter related to $T_{\rm c}$ is the pnictogen height 
from the Fe-layer ($h_{pn}$)~\cite{Mizoguchi,Kuroki}. 
However, it is not clear yet what electronic parameter is modified 
by this angle or $h_{pn}$. 
Although some theories suggest that the antiferromagnetic (AF) fluctuation 
plays an important role for the appearance of SC 
in the iron pnictides~\cite{Kuroki,Mazin}, 
there is no direct experimental evidence that $T_{\rm c}$ is correlated 
with the strength of AF fluctuation. 
Therefore, in order to clarify the mechanism of SC 
in this system, it is necessary to find a microscopic parameter 
that scales with $T_{\rm c}$, 
comparing various physical properties of various iron pnictides 
with different $T_{\rm c}$. 

In the present study, we focus on $R$FeP$_{1-x}$As$_x$O$_{0.9}$F$_{0.1}$, 
where $R$=La, Pr and Nd. 
One of the advantages of this system is that P and As are isovalent elements 
and thus a carrier number is kept constant in principle. 
The change in physical properties with $x$ is considered 
to be induced by a structural change due to chemical pressure. 
The second advantage is that we can cover a wide range of $T_{\rm c}$ 
from $\sim$3 K to $\sim$50 K 
by changing $x$. 
This helps us to find a physical quantity that scales with $T_{\rm c}$. 
Both end materials are rather well investigated. 
$R$FeAsO$_{1-y}$F$_y$ becomes an AF metal 
when F is not doped~\cite{Kamihara,Huang,Rotundu,Ren}. 
With increasing $y$, the AF order is suppressed 
and the SC emerges above $y$$\sim$0.08. 
Therefore, the end material in the present study, 
$R$FeAsO$_{0.9}$F$_{0.1}$ shows SC 
at low temperatures, but has large AF fluctuation. 
In contrast, 
the other end material $R$FePO$_{1-y}$F$_y$ are 
superconducting ~\cite{Baumbach,Suzuki}. 
Even without F-doping, it is also superconducting below 
$\sim$ 4 K, and shows 
a paramagnetic metallic behavior in the normal state. 
Therefore, the AF fluctuation is expected 
to be controllable by changing $x$ in $R$FeP$_{1-x}$As$_x$O$_{0.9}$F$_{0.1}$. 

\begin{figure}[hb]
\begin{center}
\includegraphics[width=70mm]{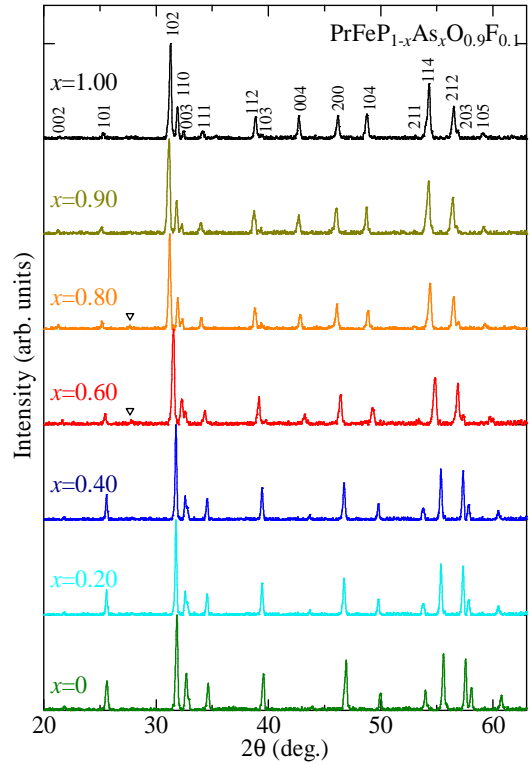}
\end{center}
\caption{(Color online) Powder X-ray diffraction patterns 
for PrFeP$_{1-x}$As$_x$O$_{0.9}$F$_{0.1}$ with various $x$s. 
Almost all the diffraction peaks are indexed 
assuming the tetragonal structure with the $P$4/$nmm$ symmetry. 
The peaks indicated by triangles are due to impurities. 
}
\label{fig1}
\end{figure}

\begin{figure*}[tb]
\begin{center}
\includegraphics[width=140mm]{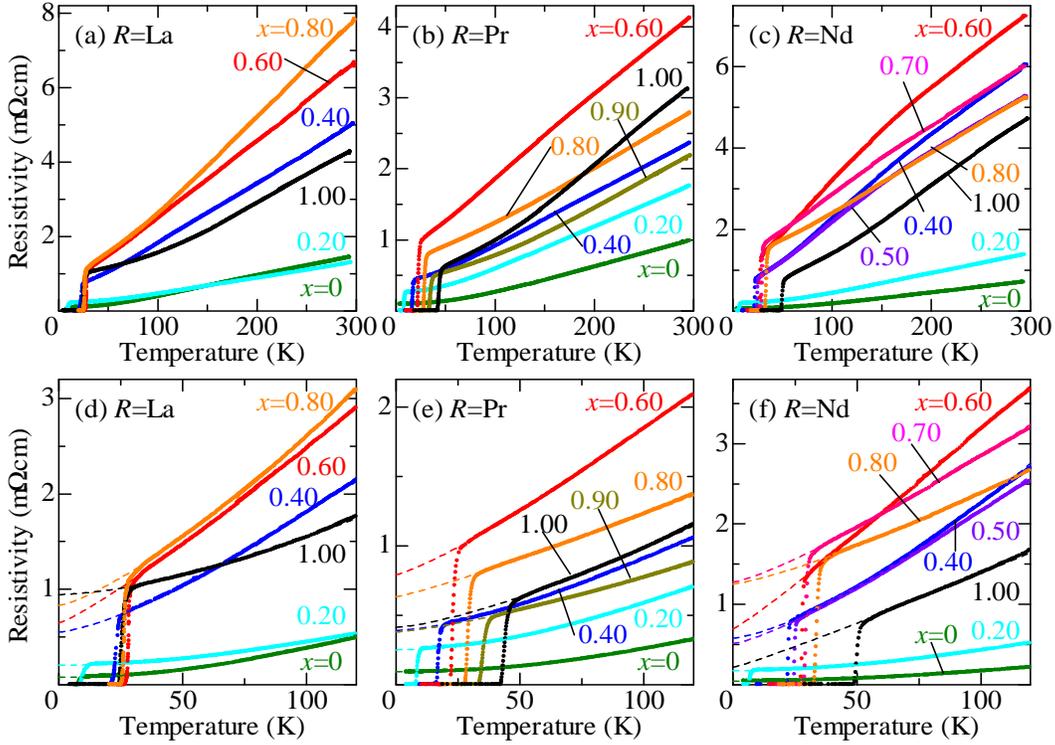}
\end{center}
\caption{(Color online) Temperature dependence of 
resistivity below room temperature ((a)-(c)) 
and below 120 K ((d)-(f)) 
for $R$FeP$_{1-x}$As$_x$O$_{0.9}$F$_{0.1}$ ($R$=La, Pr and Nd) 
with various $x$s, respectively.
In the panels (d)-(f), the dots and the broken lines indicate 
the experimental results and the fitting curves by using 
$\rho$=$\rho_0$+$AT^n$, respectively. 
The fitting of $\rho$($T$) was performed 
between the onset $T$ of resistive transition 
($T_{\rm c}^{\rm onset}$) and 100 K. 
}
\label{fig2}
\end{figure*}

A similar study was reported in BaFe$_2$(As$_{1-x}$P$_x$)$_2$~\cite{Kasahara}, 
where the AF interaction is modified by P/As-substitution. 
With increasing $x$, AF order is suppressed 
and SC manifests itself near $x$=0.33, 
giving a quantum critical behavior. 
Although we have preliminarily investigated 
F-free $R$FeP$_{1-x}$As$_x$O which shows an AF order 
at $x$=1.0, we observed neither SC with $T_{\rm c}$ 
higher than 10 K nor any anomalous behavior 
due to a magnetic quantum criticality. 
By contrast, $R$FeP$_{1-x}$As$_x$O$_{0.9}$F$_{0.1}$ 
does not have any magnetic order, but shows 
a drastic change with $x$ in physical properties. 
In the present study, we have investigated the resistivity 
($\rho$($T$)) 
and the Hall effect in $R$FeP$_{1-x}$As$_x$O$_{0.9}$F$_{0.1}$ 
with various $T_{\rm c}$, lattice constants 
and presumably AF fluctuation strength 
to find a relationship among $T_{\rm c}$, 
crystal structure and electronic properties 
in iron pnictides. 

\section{Experimental procedures}

Polycrystalline 
$R$FeP$_{1-x}$As$_x$O$_{0.9}$F$_{0.1}$ 
($x$=0$\sim$1.0) 
were synthesized 
by solid state reaction. 
The mixtures of $R$As, $R$P, Fe$_2$O$_3$, Fe and FeF$_2$ 
in the stoichiometric ratio were pressed into pellets 
in a pure Ar filled glove box 
and annealed at 1100 $^{\circ }$C for 40 h in evacuated silica tubes. 
All the samples were prepared by the same careful procedure. 
The result of EDX (Energy Dispersive X-ray spectroscopy) 
indicates that the actual F concentration 
is about 0.03 $\sim$ 0.04, 
which is smaller than the nominal one. 
Since there are peaks for $R$ and Fe near the peak for F 
in the EDX spectrum, 
we could not exactly determine the actual F concentration. 
Therefore, we show the nominal F concentration (0.1) in this paper. 

The samples were characterized by powder X-ray diffraction 
using Cu $K_{\alpha}$ radiation 
at room temperature. 
In Fig. 1, we show the powder X-ray diffraction pattern for 
PrFeP$_{1-x}$As$_x$O$_{0.9}$F$_{0.1}$ as an example. 
Almost all the diffraction peaks can be assigned to the calculated Bragg peaks 
for the tetragonal $P$4/$nmm$ symmetry. 
The in-plane ($a$) and out-of-plane lattice constants ($c$) were 
obtained by the least squares fitting of the X-ray diffraction data. 
The values of $a$ and $c$ for $x$=0 and 1.0 
well agree 
with the reported data~\cite{Kamihara,Huang,Suzuki,Ren2,Qiu,Zimmer}. 
As shown in Fig. 1, the peak position of the powder X-ray diffraction 
data is systematically changed with increasing $x$, and 
both of $a$ and $c$ linearly increase with $x$, 
as indicated later in Fig. 3(a). 
This proves that solid solutions of the present system 
have been successfully prepared, 
and the actual F concentrations are almost constant 
in the whole $x$-range. 

The magnetic susceptibility was measured in a magnetic field of 10 Oe. 
The superconducting volume fractions estimated 
from the diamagnetic susceptibility at 2 K are over 80 $\%$ 
for all the samples. 
The temperature ($T$) dependence of electrical resistivity ($\rho$($T$)) 
was measured 
by a standard four-probe method from room $T$ down to 4.2 K. 
The Hall coefficient $R_{\rm H}$ was measured in magnetic fields 
up to 7 T at various $T$s.

\section{Results and discussion}

\begin{figure}[h]
\begin{center}
\includegraphics[width=70mm]{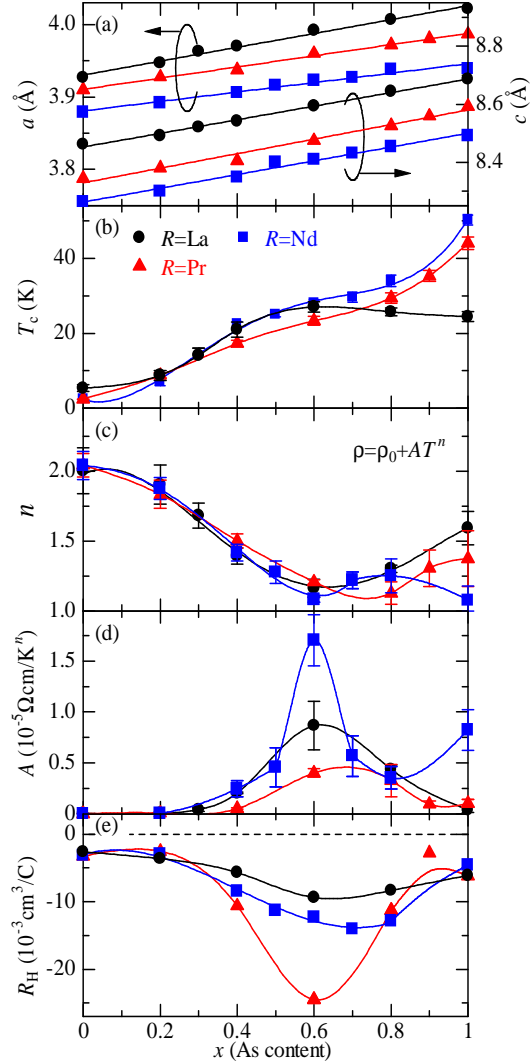}
\end{center}
\caption{(Color online) $x$ dependence 
of physical properties for $R$FeP$_{1-x}$As$_x$O$_{0.9}$F$_{0.1}$. 
(a)Lattice constants of $a$- and $c$-axes 
at room temperature. (b)Critical temperature 
$T_{\rm c}$. (c)Power $n$ of temperature in resistivity 
($\rho$=$\rho_0$+$AT^n$). (d)The coefficient $A$. 
(e)Hall coefficient ($R_{\rm H}$) at 50 K. 
The error bars for $T_{\rm c}$ in panel (b) are estimated 
from the onset and zero resistivity temperatures. 
We performed the fitting of $\rho$($T$) in the $T$-ranges 
of $T_{\rm c}^{\rm onset}<T<$80 K and 
$T_{\rm c}^{\rm onset}<T<$120 K, 
and estimated the error bars of $n$ and $A$ in the panels 
(c) and (d).
}
\label{fig3}
\end{figure}

Figures 2(a)-(c) show the $\rho$($T$) 
with various $x$s for $R$=La, Pr and Nd, respectively. 
In almost all the samples, the superconducting transitions are 
sharp enough to determine $T_{\rm c}$ from the midpoint of $T$ 
of the resistive transition. 
(In the $x$=0 samples of $R$=Pr and Nd, $T_{\rm c}<$4.2 K is defined 
by an onset transition $T$ in magnetic susceptibility.) 
In contrast to the linear $x$-dependence of $a$ and $c$ (Fig. 3(a)), 
$T_{\rm c}$ does not monotonically change with $x$ (Fig. 3(b)). 
In all the systems, $T_{\rm c}$ gradually increases with $x$ up to $x$=0.60, 
while the behavior changes above $x$=0.60. 
For $R$=La, $T_{\rm c}$ saturates at $x$$\sim$0.6 
and slightly decreases above $x$=0.6, 
while for $R$=Pr and Nd, $T_{\rm c}$ is more rapidly enhanced 
above $x$=0.80 than that for $x$$<$0.6. 

Non-monotonic $x$-dependence was also observed in $\rho$($T$). 
As shown in Figs. 2(a)-(f), $\rho$($T$) for all the samples 
exhibits a metallic behavior. 
In all the $R$-systems, the resistivity value is the lowest at $x$=0. 
With increasing $x$, the residual resistivity $\rho_0$ 
and the slope of $\rho$($T$) 
are rapidly enhanced, showing a maximum at $x$=0.60$\sim$0.80. 
Such a non-monotonic but systematic change of $\rho$($T$) with $x$ 
was observed in all the $R$-systems, which indicates 
that the observed change is intrinsic, 
but not due to a grain boundary effect. 

As shown in Figs. 2(d)-(f), the $\rho$($T$) can be expressed 
as $\rho$($T$)=$\rho_0$+$AT^n$ at low $T$s, 
where $n$ is the power of $T$ and $A$ the coefficient. 
The fitting of $\rho$($T$) was performed 
between the onset $T$ of resistive transition 
($T_{\rm c}^{\rm onset}$) and 100 K. 
Figure 3(c) shows the $x$ dependence of $n$. 
For $x$=0, $n$ is close to 2 in all the $R$-systems, 
suggesting that the end materials with $x$=0 are 
a conventional Fermi liquid. 
As $x$ increases, $n$ decreases and reaches about unity 
at $x$=0.60$\sim$0.80. 
Above $x$=0.60$\sim$0.80, $n$ slightly varies, 
but is still below 1.6. 
The $T$-linear $\rho$($T$) ($n$=1) is observed also 
in high-$T_{\rm c}$ cuprates~\cite{Takagi} 
and heavy fermion compounds~\cite{Gegenwart} 
near the quantum critical point, which suggests 
that the conduction mechanism is governed 
by strong bosonic fluctuation such as 
AF fluctuation~\cite{Moriya}.

The $x$ dependence of power $n$ indicates that the chemical pressure 
induced by the P/As substitution rapidly increases 
bosonic fluctuation with $x$ up to $x$=0.60$\sim$0.80. 
In fact, the NMR study 
in $R$=La system detected 
the strong AF fluctuation around $x$=0.6, 
while almost no AF fluctuation at $x$=1.0~\cite{Mukuda}. 
As shown in Fig. 3(d), $A$ is enhanced 
at $x$=0.60$\sim$0.80 and decreased toward $x$=1.0. 
The $A$ depends on the power $n$, and it is difficult 
to extract the physical origin only from the behavior of $A$. 
But all the $R$-systems show similar and systematic 
$x$-dependence of $A$, 
and the behavior of $A$ may be related with the enhancement of 
the AF fluctuation at $x$=0.60$\sim$0.80. 
Here we note that $\rho_0$ is also enhanced near $x$=0.6 
in all the $R$-system, which cannot be explained 
by spin fluctuation theory. 
Although it is hard to discuss absolute values of $\rho$($T$) 
for polycrystalline samples, the change of $\rho_0$ in Fig. 2 
is quite systematic and common in all the $R$-systems. 

\begin{figure*}[tb]
\begin{center}
\includegraphics[width=160mm]{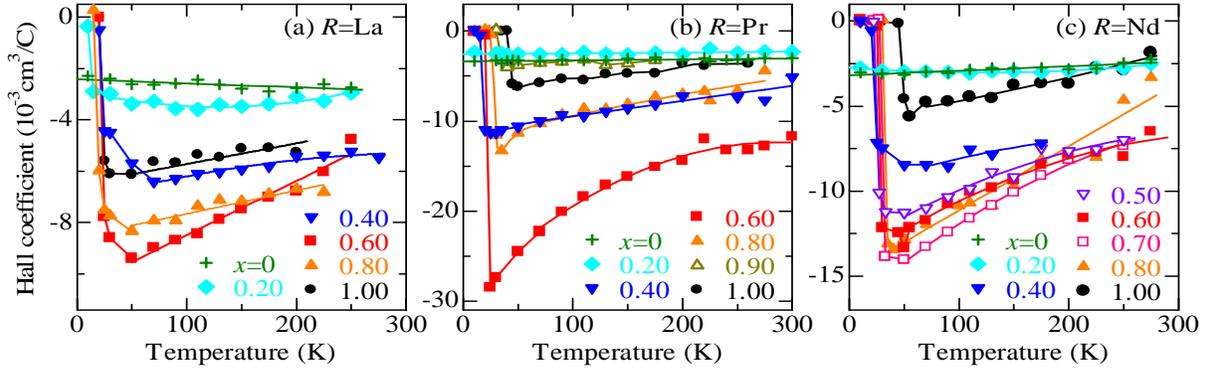}
\end{center}
\caption{(Color online) Temperature dependence of 
Hall coefficient ((a)-(c)) 
for $R$FeP$_{1-x}$As$_x$O$_{0.9}$F$_{0.1}$ ($R$=La, Pr and Nd) 
with various $x$s, respectively. 
The dots indicate the experimental results and the lines 
are guides for eyes.
}
\label{fig4}
\end{figure*}

Figures 4(a)-(c) represent $T$-dependence of Hall coefficient 
$R_{\rm H}$ with various $x$s for $R$=La, Pr and Nd, 
respectively. 
$R_{\rm H}$ at $x$=0 is almost $T$-independent 
and has a small value ($\sim2-3$ C/cm$^3$), 
while at $x$=1.0 $R_{\rm H}$ is also small 
but shows a weak $T$-dependence. 
Our new finding is that the magnitude and the $T$-dependence 
of $R_{\rm H}$ are strongly enhanced around $x$=0.60$\sim$0.80 
in all the $R$-systems. 
Above $x$=0.60$\sim$0.8, they are suppressed with $x$. 
The $x$ dependence of $R_{\rm H}$ at 50 K is plotted in Fig. 3(e). 
$R_{\rm H}$ has a broad minimum around $x$=0.60$\sim$0.8 in all the systems. 
All the non-monotonic $x$-dependences of $T_{\rm c}$, 
$n$, $A$ and $R_{\rm H}$ seen in Fig. 3 
demonstrate a critical change 
in the electronic state around $x$=0.6$\sim$0.8. 
This critical concentration $x$ may be slightly dependent 
on the $R$ element.

\begin{figure}[hb]
\begin{center}
\includegraphics[width=70mm]{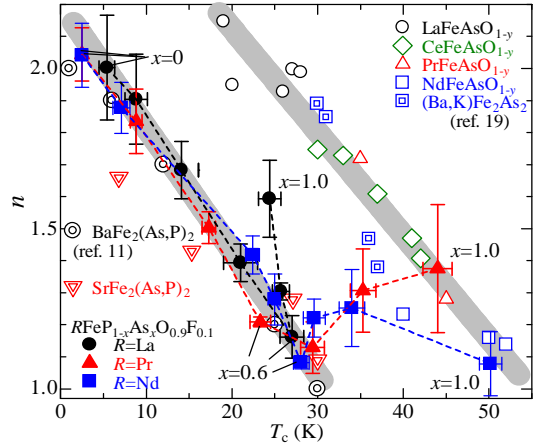}
\end{center}
\caption{(Color online) Relation between $T_{\rm c}$ 
and the power $n$ of temperature in $\rho$($T$) 
for $R$FeP$_{1-x}$As$_x$O$_{0.9}$F$_{0.1}$ ($R$=La, Pr and Nd) 
and other ion pnictides.
The closed symbols with broken lines and open ones indicate 
the present results for $R$FeP$_{1-x}$As$_x$O$_{0.9}$F$_{0.1}$ 
($R$=La, Pr and Nd) and the previous ones 
for $R$FeAsO$_{1-y}$ in ref.~\cite{Ishida}, respectively. 
The broad grey lines represent the two different correlations 
between $T_{\rm c}$ and $n$ in 
$R$FeP$_{1-x}$As$_x$O$_{0.9}$F$_{0.1}$ ($x$$<$0.6$\sim$0.8) 
and $R$FeAsO$_{1-y}$. 
The results for (Ba,K)Fe$_2$As$_2$~\cite{Ishida}, 
BaFe$_2$(As,P)$_2$~\cite{Kasahara} 
and SrFe$_2$(As,P)$_2$~\cite{Kobayashi,Kobayashi2} 
can also be plotted on these two lines.}
\label{fig5}
\end{figure}

Another piece of evidence for the electronic change 
around $x$=0.6$\sim$0.8 can be seen in the relation between $T_{\rm c}$ 
and $n$ in Fig. 5. 
The samples with $x<$0.6$\sim$0.8 show the almost linear relationship 
between $T_{\rm c}$ and $n$, which is universal for all the $R$-systems. 
The $x$=0 samples exhibit the lowest $T_{\rm c}$ and $T^2$ resistivity, 
while those with $x$=0.6$\sim$0.8 show $T_{\rm c}$$\sim$30 K 
and almost $T$-linear $\rho$($T$). 
This distinct relation suggests that what causes the $T$-linear 
$\rho$($T$) is strongly involved in the mechanism of 
high $T_{\rm c}$ SC in this system. 
A similar correlation between $T_{\rm c}$ and $n$ has been observed 
in $R$FeAsO$_{1-y}$~\cite{Ishida}, although the scaling line is 
shifted in parallel about 20 K from the present one. 

In contrast to the samples below $x$=0.6$\sim$0.8, 
the larger $x$ samples show 
no clear relation between $T_{\rm c}$ and $n$.
For $R$=Pr and Nd, 
$T_{\rm c}$ is continuously increased with $x$, 
while $n$ is almost unchanged ($n$=1$\sim$1.4) 
at $x>$0.6$\sim$0.8. 
It suggests that the $T_{\rm c}$-rising mechanism 
for $x$$>$0.6$\sim$0.80 
is different from that for $x$$<$0.6$\sim$0.80. 
It is interesting that the data point for the $x$=1.0 samples 
with $R$=Pr and Nd are on another linear correlation 
for $R$FeAsO$_{1-y}$~\cite{Ishida}, as shown in Fig. 5. 
It means that the As 100\% compounds 
of $R$FeP$_{1-x}$As$_x$O$_{0.9}$F$_{0.1}$ show 
the $T_{\rm c}$-$n$ values sitting on the right correlation line 
for higher $T_{\rm c}$, irrespective of F-content, 
while the values for $x$$<$0.6$\sim$0.80 are on the left line 
and those for $\sim$0.6$<$$x$$<$1.0 are between the two lines. 
Although the data for $x$=1.0 sample with $R$=La is not located 
on the right line in Fig. 5, 
the data points for x$>$0.6 
approach those for LaFeAsO$_{1-y}$ with increasing $x$. 
One may consider that the left $T_{\rm c}$-$n$ line 
is sifted by 20 K because of the disorder induced pair-breaking effect 
due to As/P substitution. 
However, this is unlikely because the disorder effect 
on $T_{\rm c}$ is not strong in the present system. 
It is supported by the fact that $h_{pn}$ 
determined by a precise Rietveld analysis 
for PrFeP$_{1-x}$As$_x$O$_{0.9}$F$_{0.1}$ 
exactly follow the universal 
$h_{pn}$-$T_{\rm c}$ curve~\cite{Takemori}. 

What happens at the critical As-content ($x$=0.6$\sim$0.8)? 
The theoretical calculations have predicted that the Fermi surfaces (FSs) 
of $R$FeAsO ($x$=1.0) and $R$FePO ($x$=0) are very similar 
in many aspects but do have some differences~\cite{Kuroki,Thomale}. 
Experimentally, the FSs around $\Gamma$ and M points 
of these systems have been confirmed by the angle-resolved photoemission 
spectroscopy~\cite{Lu1,Lu2,Nishi}. 
The most prominent difference predicted by theories 
is that the hole-type FS 
around ($\pi$,$\pi$,0) is missing in $R$FePO, 
while it is present in $R$FeAsO~\cite{Kuroki,Thomale}. 
Since this hole FS appears when the $d_{x^2-y^2}$ band 
crosses Fermi energy ($E_F$), it is likely that the $d_{x^2-y^2}$ band 
touches $E_F$ at $x$=0.6$\sim$0.8, which gives a critical change 
in the electronic state. 
According to this picture, the electronic properties 
for $x$$<$0.6$\sim$0.8 
should be discussed without the cylindrical hole FS 
around ($\pi$,$\pi$,0) but with two cylindrical FSs near $\Gamma$ 
and M points and one three dimensional FS near ($\pi$,$\pi$,$\pi$) 
originating from $d_{z^2}$ orbital. 
A systematic change in $n$ of $\rho$($T$) might be caused 
by the gradual enhancement of bosonic fluctuation 
due to the change of FS topology (size and shape) around $\Gamma$ 
and M points with increasing $x$. 
The maximum $T_{\rm c}$ is about 30 K in this configuration of FS. 
On the other hand, for $x$$>$0.6$\sim$0.8, the hole FS 
around ($\pi$,$\pi$,0) originating from $d_{x^2-y^2}$ orbital 
provides an additional FS nesting channel. 
This additional channel may contribute to 
a further increase in $T_{\rm c}$. 

The different lattice constants for different $R$-systems 
should give different FSs. 
Nevertheless, the critical $x$-value ($x$=0.6$\sim$0.8) 
is only a little dependent on the $R$ element. 
In these systems, the electronic structure and the FSs are 
closely dependent on the local structure around Fe ions. 
The present results indicate the P/As substitution 
linearly changes not only the lattice constants 
but also perhaps the local structure around Fe ions, 
and resultantly modifies the electronic 
state~\cite{Takemori}. 
The local structural parameter such as $h_{pn}$ 
is a more important parameter 
that determines the electronic properties. 
A relevant experimental report was made for CeFeP$_{1-x}$As$_x$O 
that the AF order in the As-rich compositions 
disappeared at $x$=0.6~\cite{Luo}. 

Next, we discuss the origin of observed anomalies 
around $x$=0.6$\sim$0.8. 
In most of the iron pnictides, 
AF phase is close to the superconducting one. 
Therefore, AF fluctuation via the FS nesting is a strong candidate 
for a pairing interaction that may also govern the transport properties. 
However, among the observed anomalies around $x$=0.6$\sim$0.8, 
the increase of $R_{\rm H}$, $\rho_0$ and $A$ 
cannot be explained by the spin fluctuation theory~\cite{Moriya}, 
and requires something others. 

Another candidate is charge fluctuation. 
We point out that similar enhancements of $R_{\rm H}$, $\rho_0$, $A$ 
and $T_{\rm c}$ together with $T$-linear $\rho$($T$) 
were observed in CeCu$_2$(Si,Ge)$_2$~\cite{Yuan,Yuan2,Seyfarth}. 
Apart from a magnetic quantum critical point 
in the pressure-$T$ phase diagram, 
this heavy fermion compound shows another critical behavior 
at a higher pressure where the $T_{\rm c}$ reaches the highest value. 
The observed anomalies were interpreted 
as a result of the rapid change of the Ce valence. 
Watanabe $et$ $al$. successfully explained these anomalies 
by the microscopic theory for valence fluctuation based 
on an extended Anderson model~\cite{Watanabe}. 
In the case of $R$FeP$_{1-x}$As$_x$(O,F), 
P/As-substitution is an isovalent substitution 
in a chemical sense. 
However, it is likely that the exchange of band energy 
with $x$ (the $d_{x^2-y^2}$ band is lifted up above $E_F$ 
and the $d_{z^2}$ band shifts down below $E_F$) 
causes valence (charge) fluctuation 
near the critical composition $x$=0.6$\sim$0.8. 
Below $x$=0.6$\sim$0.8, this charge fluctuation gradually 
increases with $x$ and causes the enhancement of $T_{\rm c}$. 

Finally, we address the issue of $T_{\rm c}$-rising mechanism. 
Although the enhancement of $R_{\rm H}$, $\rho_0$ and $A$ is the largest 
and $n$ is close to 1 near $x$=0.6$\sim$0.8, 
$T_{\rm c}$ is not a maximum at this composition 
but it further increases for larger $x$. 
This is because, as shown in Fig. 5, 
there exists another $T_{\rm c}$-$n$ line (high $T_{\rm c}$ line), 
and the data above $x$=0.6$\sim$0.8 seem to approach 
towards this line. 
The samples with $\sim$0.6$<x<$1.0 have FSs with $d_{x^2-y^2}$ 
and $d_{z^2}$ orbital characters, 
and the two $T_{\rm c}$-rising mechanisms perhaps 
by different nesting conditions 
and/or different bosonic fluctuations coexist. 
As a result, the samples with $\sim$0.6$<x<$1.0 
show a crossover behavior and their results are located 
between two $T_{\rm c}$-$n$ lines in Fig. 5. 
We also plot the data for other iron pnictides 
such as (Ba,K)Fe$_2$As$_2$~\cite{Ishida}, 
BaFe$_2$(As,P)$_2$~\cite{Kasahara} 
and SrFe$_2$(As,P)$_2$~\cite{Kobayashi,Kobayashi2} 
(A-122 system where A=Ba, Sr and K). 
At a glance, we find that all the compounds are classified 
into two groups with the two universal $T_{\rm c}$-$n$ relations, 
namely, the compounds which obey the left $T_{\rm c}$-$n$ relation 
(low $T_{\rm c}$ line) and those which obey the right one 
(high $T_{\rm c}$ line). 
On the low $T_{\rm c}$ line, $T_{\rm c}$ is enhanced with $x$ 
owing to a gradual increase of spin or charge fluctuation. 
The As/P concentration seems to be a crucial parameter 
to control the pairing interaction, 
while the lattice constant controlled by the $R$-element 
in $R$FeP$_{1-x}$As$_x$O$_{0.9}$F$_{0.1}$ 
or the A-element in 122-systems 
does not play an important role. 
The maximum $T_{\rm c}$ value in this class of compounds is about 30 K. 

By contrast, all the compounds on the high $T_{\rm c}$ line 
are P-free. 
$T_{\rm c}$ varies with the lattice parameters controlled 
by the oxygen content and/or the size of $R$-element 
in $R$FeAsO$_{1-y}$ 
or the A-element in 122-systems. 
The maximum $T_{\rm c}$ reaches 55 K in this class of compounds, 
but there was reported no clear correlation 
between AF fluctuation and $T_{\rm c}$ (or $n$). 
Therefore, the $T_{\rm c}$-rising mechanism along this line is unclear. 
It is also unknown why the high $T_{\rm c}$ line is shifted by 20 K 
from the low $T_{\rm c}$ line. 

It may be worth to note here that a nodal superconducting gap 
was reported for many compounds on the low $T_{\rm c}$ line, 
while a full gap for the compounds on the high $T_{\rm c}$ line. 
A qualitative difference in the FSs as revealed 
in the present study could contribute to this symmetry difference 
of the superconducting gap. 
All these facts related to the two $T_{\rm c}$-$n$ lines 
suggest that there exist two different $T_{\rm c}$-rising mechanisms 
in the iron pnictide superconductors 
and in some cases the two may act additively. 
\section{Conclusion}
In summary, we have clarified the relation 
between $T_{\rm c}$ and the transport properties 
by changing the As/P ratio in $R$FeP$_{1-x}$As$_x$O$_{0.9}$F$_{0.1}$ 
with $R$=La, Pr and Nd. 
It has been revealed that there are two distinct regions of $x$. 
In the low $x$-region ($x$$<$0.6$\sim$0.8), $T_{\rm c}$ linearly increases 
from 3 K to 30 K with decreasing the power $n$ in 
$\rho$($T$)=$\rho_0$+$AT^n$ from 2 (at $x$=0) to 1 (around $x$=0.6$\sim$0.8). 
This strongly suggests that some bosonic fluctuation 
is a primary factor to enhance $T_{\rm c}$. 
The universal $T_{\rm c}$-$n$ relation holds for all the 
$R$FeP$_{1-x}$As$_x$O$_{0.9}$F$_{0.1}$ 
with $x$$<$0.6$\sim$0.8 in the present study 
as well as BaFe$_2$(As,P)$_2$ and SrFe$_2$(As,P)$_2$. 
In addition to the $T$-linear $\rho$($T$), $R_{\rm H}$, $\rho_0$ and $A$ 
are strongly enhanced near $x$=0.6$\sim$0.8, 
suggesting some critical change of the electronic state. 
In the high $x$-region ($x$$>$0.6$\sim$0.8), on the other hand, 
$T_{\rm c}$ becomes strongly $R$-dependent and further increases with $x$, 
but shows no clear correlation with $n$. 
The compounds with $x$$>$0.6$\sim$0.8 seem to approach another universal 
$T_{\rm c}$-$n$ relation which holds 
for $R$FeAsO$_{1-y}$ and (Ba,K)Fe$_2$As$_2$. 
The presence of two distinct $T_{\rm c}$-$n$ relations 
could be the evidence that there are two $T_{\rm c}$-rising mechanisms 
in the iron pnictides.

\begin{acknowledgment}
We thank H. Mukuda and K. Miyake for helpful discussion and 
N. Chikumoto for support of EDX measurement. 
This work was supported by JST, TRIP and IRON-SEA. 
\end{acknowledgment}

\end{document}